\documentclass[runningheads,a4paper]{llncs}
\usepackage{amsmath, amssymb}
\usepackage{graphicx}
\usepackage{algorithmic}
\usepackage{algorithm}

\newcommand{\E}{\mathrm{E}}

\newcommand{\bR}{\mathbb{R}}

\def\argmin{\mathop{\rm argmin}}

\newtheorem{observation}[theorem]{Observation}
\newcommand\ignore[1]{}
\newtheorem{fact}{Fact}

\begin{document}
\title{Differential Privacy and the Fat-Shattering Dimension of Linear Queries}
\author{Aaron  Roth\thanks{This work has been supported in part by an NSF Graduate Research Fellowship.}}
\institute{Computer Science Department \\ Carnegie Mellon University \\ and \\ Microsoft Research New England}

\maketitle
\abstract{In this paper, we consider the task of answering linear queries under the constraint of differential privacy. This is a general and well-studied class of queries that captures other commonly studied classes, including predicate queries and histogram queries. We show that the accuracy to which a set of linear queries can be answered is closely related to its \emph{fat-shattering dimension}, a property that characterizes the learnability of real-valued functions in the agnostic-learning setting.}

\section{Introduction}
The administrator of a database consisting of sensitive, but valuable information faces two conflicting objectives. Because the data is valuable, she would like to make statistical information about it available to the public. However, because the data is sensitive, she must take care not to release information that exposes the data of any particular individual in the data set. The central question in the field of \emph{private data analysis} is how these two objectives can be traded off, and more specifically, how many queries of what type can be answered to given degrees of accuracy, while still preserving privacy.

Recent work on \emph{differential privacy} provides a mathematical framework to reason about such questions. Informally, a probabilistic function $f$ from a database $D$ to some range $\mathcal{R}$ is $\alpha$-differentially private if adding or removing a single individual from the dataset does not change the probability that $f(D) = r$ for any outcome $r \in \mathcal{R}$ by more than an $e^\alpha$ factor. The intuition behind this definition is that an individual's privacy should not be considered to have been violated by some event $r$, if $r$ would have been almost as likely to occur even without the individual's data.

In this paper, we consider databases $D$ which are real valued vectors, and the class of queries that we consider correspond to linear combinations of the entries of $D$. Formally, we consider databases $D \in \bR^n_+$, and queries of the form $q \in [0,1]^{n}$. The answer to query $q$ on database $D$ is simply the dot-product of the two vectors: $q(D) = q\cdot D$. This model has previously been considered (\cite{DN03,DMT07,DY08,HT10}), and generalizes the class of \emph{count queries} or \emph{predicate queries}, which has also been well studied (\cite{DMNS06,BLR08,DNRRV09,RR10,UV10}).

The \emph{fat-shattering dimension} (FSD) of a class of real-valued functions $C$ over some domain is a generalization of the Vapnik-Chervonenkis dimension, and characterizes a distribution-free convergence property of the mean value of each $f \in C$ to its expectation. The fat-shattering dimension of a class of functions $C$ is known to characterize the sample complexity necessary to PAC learn $C$ in the agnostic framework \cite{ABCH97,BLW94}: that is, ignoring computation, the sample complexity that is both necessary and sufficient to learn $C$ in the agnostic framework is polynomially related to the fat-shattering dimension of $C$.

Our main result is a similar information theoretic characterization of the magnitude of the noise that must be added to the answer to each query in some class $C$ in terms of the fat-shattering dimension of $C$, FSD$(C)$. We show polynomially related information theoretic upper and lower bounds on the noise that must be added to each query in $C$ in terms of FSD$(C)$. This generalizes the results of \cite{BLR08} to linear queries, and to our knowledge gives the first analysis of generic linear queries using some parameter other than their cardinality. This yields the first mechanism capable of answering a possibly infinite set of generic linear queries, and the first non-trivial lower bound for infinite classes of non-boolean linear queries. As a consequence, we extend results of Kasiviswanathan et al. and Blum et al. \cite{KLNRS09,BLR08} relating the sample complexity necessary for agnostic PAC learning and private agnostic PAC learning from classes of boolean valued functions to classes of real valued functions.

\subsection{Related Work and Our Results}
Dinur and Nissim studied the special case of linear queries for which both the database and the query are elements of the boolean hypercube $\{0,1\}^n$ \cite{DN03}. Even in this special case, they showed that there cannot be any private mechanism that answers $n$ queries with error $o(\sqrt{n})$, because an adversary could use any such mechanism to reconstruct a $1-o(1)$ fraction of the original database, a condition which they called \emph{blatant non-privacy}. This result was strengthened by several subsequent papers \cite{DMT07,DY08,KRSU10}.

Beimel et al. consider the class of basis vectors on the boolean hypercube, and show that even though this class has a constant VC-dimension (and hence fat-shattering dimension), it requires a superconstant number of samples for useful private release \cite{BKN10}. Specifically, they show that the $\log n$ factor which appears in the upper bound in this paper and in \cite{BLR08}, but not in the lower bound, is in fact necessary in some cases.

Dwork et al. gave the original definition of differential privacy, as well as the Laplace mechanism, which is capable of answering any $k$ ``low sensitivity'' queries (including linear queries) up to error $O(k)$. A more refined analysis of the relationship between the laplace mechanism and function sensitivity was later given by \cite{NRS07}.

In a different setting, Blum Ligett and Roth considered the question of answering \emph{predicate queries} over a database drawn from some domain $X$ \cite{BLR08}. This can be viewed as a special case of linear queries in which the queries are restricted to lie on the boolean hypercube, and the database must be integer valued: $D \in \mathbb{Z}_+^n$. They give a mechanism for answering every query in some class $C$ with noise that depends linearly on the VC-dimension of the class of queries. This is a quantity that is at most $\log |C|$ for finite classes $C$, and can be finite even for \emph{infinite} classes. Roth and Roughgarden later gave a mechanism which achieved similar bounds in the online model, in which the mechanism does not know the set of queries that must be answered ahead of time, and instead must answer them as they arrive \cite{RR10}. We generalize the technique of \cite{BLR08,RR10} to apply to general linear queries. VC-dimension is no longer an appropriate measure of query complexity in this setting, but we show that a quantity known as Fat-Shattering dimension plays an analogous role.

Dwork et al. \cite{DNRRV09} also gave upper and lower bounds for predicate queries, which are incomparable to the bounds of \cite{BLR08,RR10} (and those presented in this paper). The upper bounds of \cite{DNRRV09} are for an approximate form of differential privacy, and have a better dependence on $\alpha$, but a worse dependence on $k$. Their lower bounds are computational, whereas the lower bounds presented in this paper are information theoretic.

Hardt and Talwar \cite{HT10} give matching upper and lower bounds on the noise that must be added for $\alpha$-differential privacy when answering $k \leq n$ linear queries of roughly $\Theta(\frac{\sqrt{k}\log (n/k)}{\alpha})$. In contrast, we prove bounds in terms of different parameters, and can handle arbitrarily (even infinitely) large values of $k$. For finite sets of $k$ queries, our mechanism adds noise roughly \\ $O\left(||D||_1^{2/3}\cdot\left(\frac{\log k\log n}{\alpha}\right)^{1/3}\right)$. Note that for some settings of the parameters, this is significantly less noise than the bounds of \cite{HT10}: specifically, for $k \geq \Omega(||D||_1^{4/3})$. To achieve low \emph{relative} error $\eta$ (i.e. error $\epsilon = \eta ||D||_1$), our mechanism requires only that $||D||_1$ be poly-logarithmic in $k$, rather than polynomial in $k$. For infinite classes of queries $|C|$, the $\log k$ in our bound can be replaced with the fat shattering dimension of the class $C$. We also show a lower bound in terms of the fat shattering dimension of the class $C$, which is the first non-trivial lower bound for infinite classes of non-boolean linear queries.
\section{Preliminaries}
A database is some vector $D \in \bR_+^n$, and a query is some vector $q \in [0,1]^n$. We write that the evaluation of $q$ on $D$ is $q(D) = q\cdot D$. We write $||D||_1 = \sum_{i=1}^nD_i$ to denote the $\ell_1$ norm of $D$, and note that for any query $q$, $q(D) \in [0,||D||_1]$. We let $C$ denote a (possibly infinite) class of queries. We are interested in mechanisms that are able to provide answers $a_i$ for each $q_i \in C$ so that the maximum error, defined to be $\max_{i \in C} |q_i(D) - a_i|$ is as small as possible. Without loss of generality, we restrict our attention to mechanisms which actually output some synthetic database: mechanisms with range $\mathcal{R} = \bR_+^n$. That is, if our mechanism outputs some synthetic database $D'$, we take $a_i$ to be $q_i(D')$ for each $i$.\footnote{This is without loss of generality, because given a different representation for each answer $a_i$ to error $\epsilon$, it is possible to compute a synthetic database $D'$ with error at most $2\epsilon$ using the linear program of \cite{DNRRV09}.}

We formalize our notion of utility and relative utility for a randomized mechanism $M$:
\begin{definition}[Usefulness and Relative Usefulness]
A mechanism $M:\bR_+^n\rightarrow\bR_+^n$ is $(\epsilon,\delta$)-useful with respect to a class of queries $C$ if with probability at least $1-\delta$ (over the internal coins of the mechanism), it outputs a synthetic database $D'$ such that:
$$\sup_{q_i \in C} |q_i(D) - q_i(D')| \leq \epsilon$$
For $0 < \eta \leq 1$, $M$ is $(\eta,\delta)$-relatively useful with respect to $C$ for databases of size $s$ if it is $(\eta ||D||_1,\delta)$-useful with respect to $C$ for all input databases $D$ with $||D||_1 \geq s$.
\end{definition}
That is, useful mechanisms should have low error for each query in $C$. We now define differential privacy:
\begin{definition}[Differential Privacy \cite{DMNS06}]
A mechanism $M:\bR_+^n\rightarrow\bR_+^n$ is $\alpha$-differentially private, if for any two databases $D_1,D_2$ such that $||D_1-D_2||_1 \leq 1$, and for any $S \subseteq \bR_+^n$:
$$\Pr[M(D_1) \in S] \leq e^\alpha\Pr[M(D_2) \in S]$$
\end{definition}
The standard notion of differential privacy need only hold for mechanisms defined over integer valued databases $D_1,D_2 \in \mathbb{N}^n$, which is a weaker condition. Our upper bounds will hold for the stronger notion of differential privacy, and our lower bounds for the weaker notion. A useful observation is that arbitrary (database independent) functions of differentially private mechanisms are also differentially private:
\begin{fact}
\label{fact:compose}
If $M:\bR_+^n\rightarrow\bR_+^n$ is $\alpha$-differentially private, and if $f:\bR_+^n\rightarrow\bR_+^n$ is a (possibly randomized) function, then $f(M)$ is $\alpha$-differentially private.
\end{fact}


\subsection{Fat Shattering Dimension}
\emph{Fat-shattering}-dimension is a combinatorial property describing classes of functions of the form $f:X\rightarrow [0,1]$ for some domain $X$. It is a generalization of the Vapnik-Chervonenkis-dimension, which is a property only of classes of boolean valued functions of the form $f:X\rightarrow \{0,1\}$. In this section, we generalize these concepts slightly to classes of linear queries, where we view our linear queries as linear combinations of functions $f:X\rightarrow [0,1]$, where we let $X$ be the set of standard basis vectors of $\bR^n$.

Let $B = \{e_i\}_{i=1}^n$ denote the set of $n$ standard basis vectors of $\bR^n$ ($e_i$ is the vector with a 1 in the $i$'th coordinate, and a 0 in all other coordinates). For any $S \subseteq B$ of size $|S| = d$, we say that $S$ is $\gamma$-shattered by $C$ if there exists a vector $r \in [0,1]^d$ such that for every $b \in \{0,1\}^d$, there exists a query $q_b \in C$ such that for each $e_i \in S$:
$$q_b(e_i) \left\{
             \begin{array}{ll}
               \geq r_i + \gamma, & \hbox{if $b_i = 1$;} \\
               \leq r_i - \gamma, & \hbox{if $b_i = 0$.}
             \end{array}
           \right.$$
Note that since the range of each query is $[0,1]$, $\gamma$ can range from $0$ to $1/2$.
\begin{definition}[Fat Shattering Dimension \cite{BLW94,KS94}]
The $\gamma$-fat-shattering dimension of a class of linear queries $C$ is:
$$\textrm{FSD}_\gamma(C) = \max \{d \in \mathbb{N} : C\ \gamma-\textrm{shatters some } S\subseteq B\ \textrm{with }|S| = d\}$$
\end{definition}
In the special case when $\gamma = r_i = 1/2$ for all $i$, note that the fat shattering dimension of a class of boolean valued functions is equal to its VC-dimension.

For finite classes $C$, we will let $k = |C|$ denote the cardinality of $C$. The following observation follows immediately from the definition of fat-shattering dimension:
\begin{observation}
For finite classes $C$, $\textrm{FSD}_\gamma(C) \leq \log k$ for all $\gamma > 0$, where $k = |C|$.
\end{observation}

%
%
\section{Lower Bound}
In this section, we show that any $\alpha$-differentially private mechanism that answers every linear query in some class $C$ must add noise at least linear in the fat-shattering dimension of $C$ at any scale. The bound that we prove in this section is in terms of the privacy parameter $\alpha$ and the fat shattering dimension of the class. It differs from the upper bound proved in the next section by several important parameters, which include a $\log n$ term and a term depending on the size of the database. Beimel et al. \cite{BKN10} have shown that the $\log n$ term in the upper bound is necessary in some contexts. The database that we construct in our lower bound is of size $O(\gamma \cdot\textrm{FSD}_\gamma(C))$. Therefore, in order to prove a nontrivial lower bound on the \emph{relative} error achievable by a private mechanism, it would be necessary to remove a factor of $\gamma$ from our current bound. This is possible in the context of VC-dimension, and we conjecture that it should also be possible for a bound in terms of fat-shattering dimension, and is merely a limitation of our techniques as present.  The problem of proving a tight lower bound encapsulating all of the relevant parameters remains an interesting open question. We now proceed with the lower bound:

\begin{theorem}
For any $\delta$ bounded away from $1$ by a constant, let $M$ be a mechanism $M$ that is $(\epsilon,\delta)$ useful with with respect to some class of linear queries $C$. If $M$ preserves $\alpha$-differential privacy, then
$$\epsilon \geq \Omega\left(\sup_{0 < \gamma \leq 1/2} \frac{\gamma^2\cdot \textrm{FSD}_\gamma(C)}{e^\alpha}\right)$$
\end{theorem}
We begin with some preliminaries which allow us to prove some useful lemmas:

Given some class of linear queries $C$ and any $\gamma > 0$, let $S \subseteq B$ be a collection of basis-vectors of size $\textrm{FSD}_\gamma(C)$ that are $\gamma$-shattered by $C$, and let $r \in [0,1]^{\textrm{FSD}_\gamma(C)}$ be the corresponding vector as in the definition of fat-shattering dimension.  We now partition $S$ into $1/\gamma$ pieces. For each $j \in \{1,\ldots,1/\gamma\}$, let:
$$S^j = \{e_i \in S : (j-1)\cdot\gamma < r_i \leq j\cdot\gamma\}$$
Since the sets $\{S_j\}$ partition $S$, By the pigeon-hole principle, there exists some $j^*$ such that $|S^{j^*}| \geq \gamma\cdot |S| = \gamma \cdot\textrm{FSD}_\gamma(C)$. Let $d = |S^{j^*}|$.

We consider subsets $T \subset S^{j^*}$ of size $|T| = d/2$. For each such subset, we consider the database $D_T = \sum_{e_i \in T}e_i$. Let $b^T \in \{0,1\}^{d}$ be the vector guaranteed by the definition of fat shattering dimension such that:
$$b^T_i = \left\{
          \begin{array}{ll}
            1, & \hbox{$e_i \in T$;} \\
            0, & \hbox{otherwise.}
          \end{array}
        \right.$$
Let $q_{T} \in C$ be the query that corresponds to $b^T$ as in the definition of fat shattering dimension, and let $C_{S^{j^*}} = \{q_{T} : T \subseteq S^{j^*}, |T| = d/2\}$.

We first show that each function $q_{T}$ takes its highest value on $D_T$ and cannot take large values on databases $D_{T'}$ for sets $T'$ that differ significantly from $T$.
\begin{lemma}
\label{lbobserve}
For all $q_{T} \in C_{S^{j^*}}$ and for all $T' \subseteq S^{j^*}$ with $|T'| = d/2$:
$$q_{T}(D_T) - q_{T}(D_{T'}) \geq \frac{\gamma}{2}\cdot|T \bigtriangleup T'|$$
\end{lemma}
\begin{proof}
\begin{eqnarray*}
q_T(D_T)-q_T(D_{T'}) &=& \sum_{e_i \in T}q_T(e_i) - \sum_{e_i \in T'}q_T(e_i) \\
&=& \left(\sum_{e_i \in T \cap T'}q_T(e_i) - q_T(e_i)\right) + \sum_{e_i \in T \setminus T'} q_T(e_i) \\
 & & - \sum_{e_i \in T'\setminus T}q_T(e_i) \\
&\geq& \left(\sum_{e_i \in T \setminus T'} r_i + \gamma\right) - \left(\sum_{e_i \in T'\setminus T}r_i - \gamma\right) \\
&\geq& 2\gamma\cdot |T \setminus T'| - \left(\max_{i \in T'\setminus T}r_i - \min_{i \in T\setminus T'}r_i\right)\cdot |T \setminus T'| \\
&\geq& \gamma \cdot |T \setminus T'|
\end{eqnarray*}
where the last inequality follows from the fact that $T, T' \subset S^{j^*}$ which was constructed such that:
$$\left(\max_{i \in S^{j^*}}r_i - \min_{i \in S^{j^*}}r_i\right) \leq \gamma$$
holds. Observing that $|T \bigtriangleup T'| = 2|T \setminus T'|$ completes the proof.
\end{proof}
With this lemma, we are ready to prove the main technical lemma for our lower bound:
\begin{lemma}
\label{lemma:lb}
For any $\delta$ bounded away from $1$ by a constant, let $M$ be an $(\epsilon,\delta)$-useful mechanism with respect to class $C$. Given as input $M(D_T)$, where $D_T$ is an unknown private database for some $T \subseteq S^{j^*}$ with $|T| = d/2$, with constant probability $1-\delta$, there is a procedure to reconstruct a new database $D_{T^*}$ such that $|T \bigtriangleup T^*| \leq \frac{4\epsilon}{\gamma}$.
\end{lemma}
\begin{proof}
Suppose that mechanism $M$ is $(\epsilon,\delta)$ useful with respect to $C$ for some constant $\delta$ bounded away from 1. Then by definition, with constant probability, given input $D_T$, it outputs some database $D'$ such that for all $q_i \in C$, $|q_i(D_T) - q_i(D')| \leq \epsilon$. For each $T' \subseteq S^{j^*}$ with $|T'| = d/2$ let:
$$v(T') = q_{T'}(D_{T'}) - q_{T'}(D')$$
and let $T^* = \argmin_{T'} v(T')$.
Therefore, we have:
\begin{equation}
\label{eq:1}
v(T^*) \leq v(T) = q_T(D_T) - q_T(D') \leq \epsilon
\end{equation}
where the last inequality follows from the usefulness of the mechanism. We also have:
\begin{eqnarray*}
v(T^*) &=& q_{T^*}(D_{T^*})-q_{T^*}(D') \\
&\geq& q_{T^*}(D_{T^*})-q_{T^*}(D_T) - \epsilon \\
&\geq& \frac{\gamma}{2}\cdot |T \bigtriangleup T^*| - \epsilon
\end{eqnarray*}
where the first inequality follows from the usefulness of the mechanism, and the second inequality follows from lemma \ref{lbobserve}.  Combining this with equation \ref{eq:1}, we get:
$$|T \bigtriangleup T^*| \leq \frac{4\epsilon}{\gamma}$$
\end{proof}
We are now ready to prove the lower bound:

\begin{proof}[Proof of Theorem]
Let $T \subset S^{j^*}$ with $|T| = d/2$ be some randomly selected subset. Let $D_T = \sum_{e_i \in T} e_i$ be the corresponding database. By lemma \ref{lemma:lb}, given $M(D_T,\epsilon)$, with probability $1-\delta$ there is a procedure $P$ to reconstruct a database $D_{T^*}$ such that $|T \bigtriangleup T^*| \leq 4\epsilon/\gamma$. Throughout the rest of the argument, we assume that this event occurs. Let $x \in T$ be an element selected from $T$ uniformly at random, and let $y \in S\setminus T$ be an element selected from $S \setminus T$ uniformly at random. Let $T' = T\setminus {x} \cup \{y\}$. Observe that:
$$\Pr[x \in P(M(D_T,\epsilon))] \geq \frac{d/2 - 2\epsilon/\gamma}{d/2} = 1 - \frac{4\epsilon}{\gamma \cdot d}$$
$$\Pr[x \in P(M(D_{T'},\epsilon))] \leq \frac{2\epsilon/\gamma}{d/2} = \frac{4\epsilon}{\gamma \cdot d}$$
Since $||D_T - D_{T'}||_1 \leq 2$, we have by the definition of $\alpha$-differential privacy and fact \ref{fact:compose}:
\begin{eqnarray*}
e^\alpha &\geq& \frac{\Pr[x \in P(M(D_T,\epsilon))]}{\Pr[x \in P(M(D_{T'},\epsilon))]} \\
&\geq& \frac{1 - \frac{4\epsilon}{\gamma \cdot d}}{\frac{4\epsilon}{\gamma\cdot d}} \\
&=& \frac{\gamma\cdot d}{4\epsilon}-1
\end{eqnarray*}
Solving for $\epsilon$, we find that:
$$\epsilon \geq \Omega\left(\frac{\gamma \cdot d}{e^\alpha}\right)$$
Since this holds for all choices of $\gamma$, the claim follows from the fact that $d \geq \gamma \textrm{FSD}_\gamma(C)$.
\end{proof}
\section{Upper Bound}
We now show that (ignoring the other important parameters), it is sufficient to add noise linear in the fat shattering dimension of $C$ to simultaneously guarantee usefulness with respect to $C$ and differential privacy. Unlike our lower bound which was not quite strong enough to state in terms of relative error, our upper bound is most naturally stated as a bound on relative error.

We make use of a theorem of Bartlett and Long \cite{BL95} (improving a bound of Alon et al. \cite{ABCH97}) concerning the rate of convergence of uniform Glivenko-Cantelli classes with respect to their fat-shattering dimension.
\begin{theorem}[\cite{BL95} Theorem 9]
\label{lem:uniformConvergence}
Let $C$ be a class of functions from some domain $X$ into $[0,1]$. Then for all distributions $\mathbb{P}$ over $X$ and for all $\eta,\delta \geq 0$:
$$\Pr\left[\sup_{f \in C}\left|\frac{1}{m}\sum_{i=1}^m f(x_i) - \E_{x\sim \mathbb{P}}[f(x)]\right| \geq \eta\right] \leq \delta$$
where $\{x_i\}_{i=1}^m$ are $m$ independent draws from $\mathbb{P}$ and
$$m = O\left(\frac{1}{\eta^2}\left(d_{\eta/5}\ln^2\frac{1}{\eta} + \ln \frac{1}{\delta}\right)\right)$$
where $d_{\eta/5} = \textrm{FSD}_{\eta/5}(C)$.
\end{theorem}
We use this theorem to prove the following useful corollary:
\begin{corollary}
\label{cor:important}
Let $C$ be a class of linear functions with coefficients in $[0,1]$ from $\bR_+^n$ to $\bR$. For any database $D \in \bR_+^n$, there is a database $D' \in \mathbb{N}^n$ with
$$||D'||_1 = O\left(\frac{d_{\eta/5}}{\eta^2}\cdot \log^2\left(\frac{1}{\eta}\right)\right)$$
such that for each $q \in C$,
$$\left|q(D)-\frac{||D||_1}{||D'||_1}q(D')\right| \leq \eta ||D||_1$$ where $d_{\eta/5} = \textrm{FSD}_{\eta/5}(C)$.
\end{corollary}
\begin{proof}
Let $B = \{e_i\}_{i=1}^n$ denote the set of $n$ standard basis vectors over $\bR^n$. Let $\mathbb{P}_D$ be the probability distribution over $B$ that places probability $D_i/||D||_1$ on $e_i$. Note that for any $q \in C$:
$$\E_{e_i\sim\mathbb{P}_D}[q(e_i)] = \sum_{i=1}^n \frac{D_i}{||D||_1}q(e_i) = \frac{1}{||D||_1}\sum_{i=1}^n q(D_ie_i) = \frac{q(D)}{||D||_1}$$
Let $x_1,\ldots,x_m$ be $m = O\left(\frac{1}{\eta^2}\left(d_{\eta/5}\ln^2\frac{1}{\eta} + \ln 2\right)\right)$ independent draws from $\mathbb{P}_D$, and let $D' = \sum_{i=1}^mx_i$. Then:
$$q(D') = \sum_{i=1}^nq(D'_ie_i) = \sum_{i=1}^m q(x_i)$$
By lemma \ref{lem:uniformConvergence}, we have:
\begin{eqnarray*}
\Pr\left[\left|\frac{q(D')}{m} - \frac{q(D)}{||D||_1}\right| \geq \eta\right] &=&  \Pr\left[\left|\frac{1}{m}\sum_{i=1}^m q(x_i) - \E_{e_i\sim\mathbb{P}_D}[q(e_i)] \right| \geq \eta\right] \\
&\leq& \frac{1}{2}
\end{eqnarray*}
In particular, there exists some $D' \in \mathbb{N}^n$ with $||D'||_1 = m$ that satisfies $\left|\frac{q(D')}{||D'||_1} - \frac{q(D)}{||D||_1}\right| \leq \eta$. Multiplying through by $||D||_1$ gives the desired bound.
\end{proof}
Armed with Corollary \ref{cor:important}, we may now proceed to instantiate the exponential mechanism over a sparse domain, analogously to the instantiation of the exponential mechanism in \cite{BLR08}.

\begin{definition}[The Exponential Mechanism \cite{MT07}]
Let $\mathcal{D}$ be some domain, and let $s:\bR_+^n\times \mathcal{D}\rightarrow \bR$ be some \emph{quality score} mapping database/domain-element pairs to some real value. Let
$$\Delta_s \geq \max_{r \in \mathcal{D}}\sup_{D_1,D_2 \in \bR^+_n : ||D_1 - D_2||_1 \leq 1}|s(D_1,r)-s(D_2,r)|$$
be an upper bound on the $\ell_1$ sensitivity of $s$. The exponential mechanism defined with respect to domain $\mathcal{D}$ and score $s$ is the probability distribution (parameterized by the private database $D$) which outputs each $r\in \mathcal{D}$ with probability proportional to:
$$r \sim \exp\left(\frac{s(D,r)\cdot \alpha}{2\Delta_s}\right)$$
\end{definition}
\begin{theorem}[McSherry and Talwar \cite{MT07}]
The exponential mechanism preserves $\alpha$-differential privacy.
\end{theorem}
We let $m =  O\left(\frac{d_{\eta/5}}{\eta^2}\cdot \log^2\left(\frac{1}{\eta}\right)\right)$, and define the domain of our instantiation of the exponential mechanism to be:
$$\mathcal{D} = \{D' \in \mathbb{N}^n : ||D'||_1 = m\}$$
We note that $|\mathcal{D}| = n^m$. Finally, we sample each $D' \in \mathcal{D}$ with probability proportional to:
\begin{equation}
\label{eq:expmech}
 D' \sim \exp\left(-\frac{\sup_{q\in C}\left|q(D)-\frac{||D||_1}{||D'||_1}\cdot q(D')\right|\alpha}{4}\right)
\end{equation}
and output the database $D_{\textrm{out}} \equiv \frac{||D||_1}{||D'||_1}\cdot D'$\footnote{If $||D||_1$ is not public knowledge, it can be estimated to small constant error using the Laplace mechanism \cite{DMNS06}, losing only additive constants in the approximation parameter $\epsilon$ and privacy parameter $\alpha$. This does not affect our results.}.
 Observe that for any two databases $D_1,D_2$ such that $||D_1 - D_2||_1 \leq 1$ we have:
\begin{eqnarray*}
\sup_{q\in C}|q(D_1)-\frac{||D_1||_1}{||D'||_1}\cdot q(D')| - \sup_{q\in C}|q(D_2)-\frac{||D_2||_1}{||D'||_1}\cdot q(D')| &\leq& \\
||D_1-D_2||_1 + \frac{|||D_1||_1 - ||D_2||_1|}{m} &\leq&  \\
1 + \frac{1}{m} & &
\end{eqnarray*}
Therefore, the distribution defined in equation \ref{eq:expmech} is a valid instantiation of the exponential mechanism, and by \cite{MT07} preserves $\alpha$-differential privacy. It remains to show that the above instantiation of the exponential mechanism yields a useful mechanism with low error. In particular, it gives us a \emph{relatively useful} mechanisms with respect to classes $C$ for databases that have size linear in the fat shattering dimension of $C$, or only logarithmic in $|C|$ for finite classes $C$. This is in contrast to the bounds of \cite{HT10} that require databases to be of size polynomial in $|C|$ before giving relatively-useful mechanisms.
\begin{theorem}
\label{thm:offline}
For any constant $\delta$ and any query class $C$, there is an $(\eta,\delta)$-relatively useful mechanism that preserves $\alpha$-differential privacy for any database of size at least:
$$||D||_1 \geq \tilde{\Omega}\left(\frac{FSD_{2\eta/5}(C)\log n}{\alpha\eta^3}\right)$$
\end{theorem}

\begin{proof}[Proof of Theorem]
Recall that the domain $\mathcal{D}$ of our instantiation of the exponential mechanism consists of all databases $D' \in \mathbb{N}^n$ with $||D'||_1 = m$ with $m =  O\left(\frac{d_{\eta/5}}{\eta^2}\cdot \log^2\left(\frac{1}{\eta}\right)\right)\}$ In particular, by corollary \ref{cor:important}, there exists a $D^* \in \mathcal{D}$ such that:
$$\left|q(D)-\frac{||D^*||_1}{||D'||_1}q(D^*)\right| \leq \eta ||D||_1$$
By the definition of our mechanism, such a $D^*$ is output with probability proportional to at least:
$$D^* \sim \exp(-\frac{\eta ||D||_1 \alpha}{4})$$
Similarly, any $D^B \in \mathcal{D}$ such that $\left|q(D)-\frac{||D^B||_1}{||D'||_1}q(D^*)\right| \geq 2 \eta ||D||_1$ is output with probability proportional to at most:
$$D^B \sim \exp(-\frac{\eta ||D||_1\alpha}{2})$$
Let $\mathcal{D}_B$ denote the set of all such $D^B$. Because $|\mathcal{D}| = n^m$, we have that:

$$\frac{\Pr[D' = D^*]}{\Pr[D' \in \mathcal{D}_B]} \geq \frac{\exp(-\frac{\eta ||D||_1 \alpha}{4})}{n^m\cdot \exp(-\frac{\eta ||D||_1\alpha}{2})} = n^{-m}\cdot \exp\left(\frac{\eta ||D||_1\alpha}{2}\right)$$
Rearranging terms, we have:
$$\Pr[D' \in \mathcal{D}_B] \leq n^m \exp\left(-\frac{\eta ||D||_1\alpha}{2}\right)$$

Solving, we find that this bad event occurs with probability at most $\delta$ for any database $D$ with:
\begin{eqnarray*}
||D||_1 &\geq& \Omega\left(\frac{m\log n}{\eta\alpha} + \log\frac{1}{\delta}\right) \\
&=& \tilde{\Omega}\left(\frac{FSD_{2\eta/5}(C)\log n}{\alpha\eta^3}\right)
\end{eqnarray*}
\end{proof}
We remark that the above mechanism is the analogue of the general release mechanism of \cite{BLR08}, and answers linear queries in the \emph{offline} setting, when all queries $C$ are known to the mechanism in advance. This is not necessary, however. In the same way as above, corollary \ref{cor:important} can also be used to generalize the Median Mechanism of Roth and Roughgarden \cite{RR10}, to achieve roughly the same bounds, but in the \emph{online} setting, in which queries arrive online, and the mechanism must privately answer queries as they arrive, without knowledge of future queries. This results in the following theorem:
\begin{theorem}
\label{thm:online}
There exists a mechanism such that for every sequence of adaptively chosen queries $q_1,q_2,\ldots$ arriving online, chosen from some (possibly infinite) set $C$ (unknown to the mechanism), the mechanism  is $(\eta,\delta)$ useful with respect to $C$ and preserves $(\alpha,\tau)$-differential privacy\footnote{This is an approximate form of differential privacy. Specifically, a mechanism $M:\bR_+^n\rightarrow\bR_+^n$ is $(\alpha,\tau)$-differentially private, if for any two databases $D_1,D_2$ such that $||D_1-D_2||_1 \leq 1$, and for any $S \subseteq \bR_+^n$:
$$\Pr[M(D_1) \in S] \leq e^\alpha\Pr[M(D_2) \in S] + \tau$$}, where $\tau$ is a negligible function of $n$, for any database $D$ with size at least:
$$||D||_1 \geq \tilde{\Omega}\left(\frac{FSD_{2\eta/5}(C)\log n}{\alpha\eta^3}\right)$$
\end{theorem}
\begin{remark}
Notice that for finite classes of linear queries, we may replace the fat shattering dimension in the bounds of both theorems \ref{thm:offline} and \ref{thm:online} with $\log |C|$ if we so choose.
\end{remark}
\section{Conclusion}
In this paper, we have generalized the techniques used by Blum Ligett and Roth, \cite{BLR08} and Roth and Roughgarden \cite{RR10} from the class of predicate queries to the more general class of \emph{linear} queries. This gives the first mechanism for answering every linear query from some class $C$ with noise that is bounded by a parameter other than the cardinality of $C$; in particular, we have given the first mechanism for answering all of the linear queries in certain \emph{infinite} classes of queries beyond predicate queries. We have shown that the relevant parameter is the Fat-Shattering dimension of the class, which is a generalization of VC-dimension to non-boolean valued queries. In particular (ignoring other parameters), it is necessary and sufficient to add noise proportional to the fat shattering dimension of $C$. Our results show, among other things, that the sample complexity needed to privately agnostically learn real valued functions is polynomially related to the sample complexity needed to non-privately agnostically learn real valued functions.

At a high level, the same technique can be applied for any class of queries, all of the answers to which can be summarized by some `small' object. It is then sufficient to instantiate the exponential mechanism only over this much smaller set of objects (rather than the set of all databases) to obtain a useful mechanism. In the case of linear queries, we have shown that the answers to many queries can be summarized by integer valued databases with small $\ell_1$ norm. An interesting future direction is to determine what types of nonlinear (but low sensitivity) queries have similar small summarizes from which useful mechanisms can be derived.

\section{Acknowledgements}
The author would like to thank Avrim Blum for many insightful discussions, and the anonymous reviewers for extremely detailed and helpful comments.

\end{document}